\documentclass[11pt]{article}
\usepackage{amsmath,amsfonts, braket}
\def \beq  {\begin{equation}}
\def \eeq  {\end{equation}}
\def \beqar {\begin{eqnarray}}
\def \eeqar {\end{eqnarray}}
\allowdisplaybreaks
\def\sqr#1#2{{\vcenter{\vbox{\hrule height.#2pt
\hbox{\vrule width.#2pt height#1pt \kern#1pt
\vrule width.#2pt}\hrule height.#2pt}}}}

\def\vx {{\vec x}}

\def\vf {{\varphi}}

\def\Tr {{\rm Tr}}

\def\vx {{\vec x}}

\def\del {\partial}

\def\D {{\cal D}}

\def\A {{\cal A}}
\def\B {{\cal B}}

\def\H {{\cal H}}

\def\G {{\cal G}}

\def\E {{\cal E}}

\def\vf {{\varphi}}

\def \H {{\cal H}}

\def\half{\textstyle{1\over 2}}

\begin{document}
%%%%%%%%%%%%%%%%%%%%%%%%%%%%%%%%%%%%%%%%%%%%%%%
%%%%%%%%%%%%%%%%%%%%%%%%%%%%%%%%%%%%%%%%%%%%%%%
%\fontfamily{pnb}\fontsize{12pt}{16pt}\selectfont
%\fontfamily{pzc}\fontsize{14pt}{16pt}\selectfont
%\fontfamily{pbk}\fontsize{12pt}{16pt}\selectfont
%\fontfamily{cmr}\fontsize{11pt}{15pt}\selectfont
\fontfamily{bch}\fontsize{11pt}{15pt}\selectfont
%\fontfamily{phv}\fontshape{ro}\fontsize{11pt}{14pt}\selectfont
%\fontfamily{ptm}\fontseries{m}\fontshape{r}\fontsize{12pt}{16pt}\selectfont
%\fontfamily{pnc}\fontseries{m}\fontshape{r}\fontsize{11pt}{15pt}\selectfont
%\fontfamily{ppl}\fontseries{m}\fontshape{r}\fontsize{11pt}{15pt}\selectfont
%\usefont{T1}{phv}{m}{it}
%%%%%%%%%%%%%%%%%%%%%%%%%%%%%%%%%%%%%%%%%%%%%%%
%%%%%%%%%%%%%%%%%%%%%%%%%%%%%%%%%%%%%%%%%%%%%%%
\def \CMP {{Commun. Math. Phys.}}
\def \PRL {{Phys. Rev. Lett.}}
\def \PL {{Phys. Lett.}}
\def \NPBProc {{Nucl. Phys. B (Proc. Suppl.)}}
\def \NP {{Nucl. Phys.}}
\def \RMP {{Rev. Mod. Phys.}}
\def \JGP {{J. Geom. Phys.}}
\def \CQG {{Class. Quant. Grav.}}
\def \MPL {{Mod. Phys. Lett.}}
\def \IJMP {{ Int. J. Mod. Phys.}}
\def \JHEP {{JHEP}}
\def \PR {{Phys. Rev.}}
\def \JMP {{J. Math. Phys.}}
\def \GRG{{Gen. Rel. Grav.}}
%%%%%%%%%%%%%%%%%%%%%%%%%%%%%%%%%%%%%%%%%%%%%%%
%%%%%%%%%%%%%%%%%%%%%%%%%%%%%%%%%%%%%%%%%%%%%%%
\begin{titlepage}
\null\vspace{-62pt} \pagestyle{empty}
\begin{center}
\rightline{~}
\rightline{~}
\vspace{1truein} {\Large\bfseries
Aspects of Boundary Conditions for Nonabelian Gauge Theories}\\
\vskip .15in
{\Large\bfseries ~}\\
\vskip .1in
{\Large\bfseries ~}\\
%%%%%%%%%%%%%%%%%%%%%%%%%%%%%%%%%%%%%%%%%%%%%%%
%%%%%%%%%%%%%%%%%%%%%%%%%%%%%%%%%%%%%%%%%%%%%%%
 {\large\sc A.P. Balachandran$^a$}, {\large\sc V.P. Nair$^b$} and {\large\sc Sachindeo Vaidya$^c$}\\
\vskip .2in
{\itshape $^a$Physics Department, Syracuse University\\
Syracuse, NY 13244-1130}\\
\vskip .1in
{\itshape $^b$Physics Department,
City College of the CUNY\\
New York, NY 10031}\\
\vskip .1in
{\itshape $^b$Centre for High Energy Physics, 
Indian Institute of Science\\
 Bangalore 560012, India}\\
 \vskip .1in
\begin{tabular}{r l}
E-mail:&\!\!\!{\fontfamily{cmtt}\fontsize{11pt}{15pt}\selectfont balachandran38@gmail.com}\\
&\!\!\!{\fontfamily{cmtt}\fontsize{11pt}{15pt}\selectfont vpnair@ccny.cuny.edu}\\
&\!\!\!{\fontfamily{cmtt}\fontsize{11pt}{15pt}\selectfont vaidya@iisc.ac.in}
\end{tabular}
\vskip 1in
%%%%%%%%%%%%%%%%%%%%%%%%%%%%%%%%%%%%%%%%%%%%%%%
%%%%%%%%%%%%%%%%%%%%%%%%%%%%%%%%%%%%%%%%%%%%%%%
\centerline{\large\bf Abstract}
\end{center}
The boundary values of the time-component of the gauge potential form externally specifiable
data characterizing a gauge theory. We point out some consequences such as reduced symmetries,
bulk currents for manifolds with disjoint boundaries and some nuances of how the charge
algebra is realized.
\end{titlepage}
%%%%%%%%%%%%%%%%%%%%%%%%%%%%%%%%%%%%%%%%%%%%%%%
%%%%%%%%%%%%%%%%%%%%%%%%%%%%%%%%%%%%%%%%%%%%%%%
\pagestyle{plain} \setcounter{page}{2}
\section{Introduction}
Boundary conditions on fields may be viewed as the idealization and simplification of 
the dynamics of these fields on the boundary of the spacetime region under consideration.
They are also necessary, from a mathematical point of view, to
make the problem well-defined with the required self-adjointness properties for
observables.
Therefore, not surprisingly, the impact of boundary conditions on physical phenomena has been the subject of  many investigations. 
In quantum field theory, the most extreme example of the importance
of boundary conditions might be topological field theories where the entire dynamics
is played out on the boundary \cite{top}. Droplets of fermions in
the quantum Hall effect, for which the effective theory is topological and 
the dynamics is reduced to that of the edge currents,
 is a physical realization of this \cite{QHE1}.
A more standard, but still
vivid, example
of boundaries and their impact on bulk phenomena is the Casimir effect, which has 
been calculated
for many different geometries and experimentally verified for many instances as well
\cite{casimir}.
Among other examples of recent research interest we mention the possibility of
edge states in gauge theories \cite{bal1}, bound states allowed by the general
von Neumann theory of self-adjoint extensions \cite{vonN}-\cite{TRG}, the role of boundary conditions and edge states for
questions of entanglement in gauge theories \cite{entang}, etc.

Closely related to the issue of boundary conditions is the asymptotic behavior of fields
and its impact on realizations of symmetry. It was recognized long ago that
Lorentz transformations cannot be unitarily implemented on charged sectors of
quantum electrodynamics due to the infrared behavior of the massless photon fields
\cite{froh}.
While this does not have any immediate impact on everyday laboratory applications 
of electrodynamics, since one can use the charge zero sector for such considerations,
this does highlight nuances of how symmetry is realized in the theory.
More recently similar effects have been analyzed in the context of gravitational
fields.
Infrared effects can also have an impact on the realization of 
symmetries other than Lorentz transformations as well.
Obstructions to the unitary realization of color transformations
or charge rotations in a nonabelian gauge theory
have been pointed out in \cite{Balachandran:2014voa}.
Again, this may not have any immediate impact on 
calculations in the realistic case of QCD, for which the asymptotic states
are expected to have zero color charge, but this does show 
the subtleties of symmetries in a gauge theory.

Another special feature is that,
unlike theories of scalar or spinor fields, 
in a gauge theory, there is a field, namely the time-component of
the gauge potential $A_0$, whose sole effect
is via its boundary values, the physical consequences of the bulk values of the field
being wiped out by gauge invariance, i.e., by
the Gauss law imposed on physical states.
The complete elimination of $A_0$, as done, for example, in 
the so-called $A_0 = 0$ gauge, is too restrictive since
the boundary value of this field is actually
gauge-invariant, in the sense of commuting
with the constraints due to gauge symmetry.
In this paper, we analyze
this situation for a nonabelian gauge theory
pointing out some interesting features and physical implications.

In the next section we set up the basic framework for our discussion.
The results are in section 3, divided into three subsections covering
manifolds with disjoint boundaries, magnetic monopoles and realizations of
the charge algebra.

%%%%%%%%%%%%%%%%%%%%%%%%%%%%%%%%%%%%%%%%%%%%%%%
%%%%%%%%%%%%%%%%%%%%%%%%%%%%%%%%%%%%%%%%%%%%%%%
\section{Basic framework}

As mentioned in the introduction, we will be focusing on the effects of nonzero 
boundary values for fields. To set the stage, we start with a brief discussion of the 
canonical set-up for gauge theories. We are primarily interested in
nonabelian gauge theories, ignoring matter for the present. Thus the Lagrangian is that
for the
Yang-Mills theory and is given by
\begin{equation}
{\cal L} = {1\over 4} \Tr \left( F_{\mu\nu} F^{ \mu\nu} \right)
= \frac{1}{2} \left (F^a_{0i}F^a_{0i}\right)   - \frac{1}{4}  \left( F^a_{ij} F^a_{ij}
\right)
\label{1}
\end{equation}
where the fields strength tensor is, as usual, given by
\beqar
F_{0i} &=& (-i T^a ) F^a_{0i} = \del_0 A_i - \del_i A_0 + [A_0, A_i],
\nonumber\\
F_{ij} &=& (-iT^a) F^a_{ij} = \del_i A_j - \del_j A_i + [A_i, A_j],
\label{2}
\eeqar
Here $A_0, A_i$ are Lie-algebra valued fields of the form
$A_0 = (-iT^a) A_0^a $, $A_i =
(-iT^a) A_i^a $;  $\{ T^a \}$ form a basis for the Lie algebra (which will be $SU(N)$
for most of our discussion); they may be taken as hermitian matrices obeying
the normalization $\Tr (T^a T^b ) = \delta^{a b}$.

As is well known, the canonical momentum for $A_i$ is given by $E_i = F_{0i}$, while the canonical
momentum for $A_0$, which we may denote by $E_0$, is zero. This is to be viewed as
a constraint imposed on the phase space made of all four fields and their momenta.
The Hamiltonian is given by
\beqar
{\cal H} &=& \int d^3x~\Bigl(- {\rm Tr}\, \left( E_i {\del_0 A}_i + E_0 \partial_0 A_0 \right) - {\cal L}\Bigr)
 \label{Ham1}\nonumber\\
 &=& \int d^3x~ \left[\frac{1}{2} (E^a_i E^a_i + B^a_i B^a_i) + E^a_i (D_i A_0 )^a+ E^a_0 \Theta^a\right] \label{Ham2}
\eeqar
Since there is freedom of adding terms proportional to the constraints
in the Hamiltonian, we can view $\del_0 A_0$ as an arbitrary Lie algebra valued function.
Put another way, there is
nothing in the theory which determines $\del_0 A_0$; we denote this quantity by
$\Theta$ in (\ref{Ham2}).
If we eliminate the constraint $E_0$ and its corresponding
conjugate constraint in the sense of Dirac's theory of constraints, 
this is equivalent to specifying $\del_0 A_0$.
 Another comment regarding the $E_i D_i A_0$ term is also in order.
It arises from the canonical definition of the Hamiltonian and we have not done any 
integration by parts. This is important since we want to analyze boundary values 
and their effects. Secondly, the energy-momentum tensor for the Yang-Mills
Lagrangian is given by
\beq
T_{\mu\nu} = 2\, {\delta S \over \delta g^{\mu\nu}}
=  \Tr \left( F_{\mu}^{~\alpha} F_{\nu\alpha} \right) - g_{\mu\nu} \,{1\over 4}
\Tr (F^2) 
\label{3}
\eeq
where $S$ is the Yang-Mills action for an arbitrary metric $g_{\mu\nu}$ and after variation 
we set the $g_{\mu\nu}$ to be the Minkowski metric.
If we now identify the Hamiltonian as the integral of $T_{00}$, then we do not
have the $E_i D_i A_0$ term. Such differences between the canonical
definition and what is defined by variation with respect to the metric
can arise for fields which transform nontrivially
under the Lorentz group. (We are referring to the Belinfante terms
which are due to spin effects.)
The Hamiltonian we are using in (\ref{Ham1}) is the standard canonical Hamiltonian.

We now define the operator
\begin{equation}
Q(\mu) = \int d^3 x \, E^a_i (D_i \mu)^a  = - \int d^3 x \, {\rm Tr} \, (E_i D_i \mu)
\label{4}
\end{equation}
where $\mu =  (-i T^a) \mu^a $ is a test function,
with $\mu^a$ being $\mathbb{R}$-valued functions.
There are three types of test functions to be considered.
\begin{enumerate}
\item
If $\mu$ vanishes at spatial infinity, i.e., if
$\mu^a \rightarrow 0 \quad {\rm as} \quad |\vec{x}| \rightarrow \infty$, $Q(\mu )$
becomes the Gauss law operator $\G (\Lambda )$. The physical states
are selected by the condition that they are annihilated by $\G(\Lambda )$.
(We use the letter $\Lambda$ for those functions $\mu$ which vanish at
infinity.) Thus physical states obey
\beq
G(\Lambda ) \, \ket{\rm phys} =  \int d^3 x{\rm Tr} \, (E_i D_i \Lambda)\, \ket{\rm phys}
= \int d^3 x{\rm Tr} \, (- D_iE_i ) \Lambda\, \ket{\rm phys}
= 0\,.
\label{physical1}
\eeq

\item  Now consider test functions $\mu\,\, | \mkern-15mu \rightarrow 0$ as $\vert\vec{x}\vert \rightarrow \infty$,
where the symbol $| \mkern-15mu \rightarrow$ stands for 
``does not necessarily go to". The first situation is the case when
$\mu$ does not vanish at infinity, but is a constant, independent of angles,
as $\vert \vx \vert \rightarrow \infty$. We then define
\beqar
\mathcal{G}_0 &=& \left\{ {\rm Set~ of ~gauge ~transformations~} g : {\mathbb R}^3 \rightarrow
SU(N), ~{\rm such~ that} ~ g(x) \rightarrow {\rm constant,}\right.\nonumber\\
&&\hskip .1in\left. {\rm  ~not~ necessarily~} 1 ~{\rm as}~ \vert\vx \vert \rightarrow
\infty \right\}\nonumber\\
\mathcal{G}_0^\infty &=& \left\{ {\rm Set ~of ~gauge~ transformations~} g : {\mathbb R}^3 \rightarrow
SU(N), ~{\rm such ~that} ~ g(x) \rightarrow 1 \right.\nonumber\\
&&\hskip .1in \left.~{\rm as}~ \vert\vx \vert \rightarrow
\infty\right\}\nonumber
\eeqar
Thus $\G(\Lambda)$ generate the group $\mathcal{G}_0^\infty$, while
the $Q(\mu)$'s generate the group 
$\mathcal{G}_0$. The quotient
$\mathcal{G}_0/\mathcal{G}_0^\infty$ is the global group $G$, which we take to be
 $SU(N)$ for most of our discussions.

\item More generally, we can have the situation where
 the test functions $\mu \rightarrow$ angle-dependent limit, 
i.e., it is a nonconstant function on the two-sphere at spatial infinity. In this case, the
operators $Q(\mu)$ generate the so-called
Sky group \cite{sky}.

\end{enumerate}
The operators $Q(\mu )$ obey the commutation algebra
\begin{equation}
[Q(\mu_1), Q(\mu_2)]= - i\,Q([\mu_1,\mu_2]). \label{chargecomm}
\end{equation}
This is easily verified using the canonical equal-time commutation rules
\begin{equation}
[A_i^\alpha(\vec{x}), E_j^\beta(\vec{y})] = i \,\delta^{\alpha \beta} \,\delta_{ij}\,\delta^{(3)}(\vec{x} - \vec{y})
\label{CCR}
\end{equation}
For the case of $\mu$ being a constant at spatial infinity, and on states obeying the
Gauss law constraint 
(\ref{physical1}), (\ref{chargecomm}) reduces to $SU(N)$ Lie algebra.

We now briefly go back to the point made after (\ref{Ham2}) about $\del_0 A_0$.
The constraint $E_0 \approx 0$ is one of the primary constraints in the theory.
Its time-evolution can generate a secondary constraint.
Consider the time-evolution of $E_0$ smeared with a test function $\Theta $, with
$\Theta\,\, | \mkern-15mu \rightarrow 0$. The commutator with the Hamiltonian is
given by
\beq
[  \int d^3x\,\Tr(E_0 \,\Theta), {\cal H} \, ]  = 
i \, Q(\Theta)
\label{5}
\eeq
This shows that the imposition of the constraint $\int \Tr (E_0 \Theta ) \ket{\psi} = 0$ on the states would in turn require the vanishing of $Q(\Theta )$ where $\Theta\,\, | \mkern-15mu \rightarrow 0$.
In particular, this would mean that the charge carried by such states
(measured by $Q(\Theta )$) for constant $\Theta$) could be zero. Thus charged states would be
ruled out by fiat, if we impose $E_0 \approx 0$ with a $\Theta$ which does not vanish at 
infinity. In eliminating the canonical pair
$E_0, \Theta$, the boundary value of $\Theta = \del_0 A_0$ should not be eliminated
if we allow charged states. This is consistent with the fact that the boundary values
of $A_0$ and, hence, its time-derivative can be chosen freely.
They are externally chosen parameters which characterize
the theory.

If $\Theta$ in (\ref{5}) does go to zero at infinity, then we have the Gauss law operator 
on the right hand side, which should annihilate physical states.
Therefore there is no difficulty or inconsistency with requiring
$\int \Tr (E_0 \Theta ) \approx 0$ for test functions which vanish
at spatial infinity.

The conclusion is that, in general, the theory does allow for nonzero values for
$A_0^\infty$ (and its time-derivative). This is what we propose to analyze
in subsequent sections.

So far we have argued that the mathematical framework allows for a nonzero $A_0$ at
spatial infinity. We will now give two physical contexts where this does arise, before
proceeding with further analysis.

First note that $A_0 dt$ is the same in both Lorentzian and Euclidean cases as both $A_0$ and $dt$ acquire $i$'s in passage to Euclidean signature. This means that some of the
arguments about
$A_0 dt$ can be directly applied to finite temperature field theory where
we have periodicity in Euclidean time, with period $\beta =  1/T$.

Now the gluons transform according to the adjoint representation ${\rm Ad}\,G$, which for QCD is ${\rm Ad}\,(SU(3))$. The fundamental group of ${\rm Ad} G$ is
$\Pi_1({\rm Ad}\,(SU(N))) = {\mathbb Z}_N$, so there are loops of gauge transformations 
$\tilde{g}(\vx ,t)$
which as $t$ varies from $0$ to $\beta=1/T$ winds from $\mathbf{1}$ to $\mathbf{1}$ non-trivially.
(We may think of $t$ as (Euclidean) time, so that the loop is traced out as time evolves.)
 In $SU(N)$, the image of this set of tranformations
 goes from $g(\vx, 0)=\mathbf{1}$, to $g(\vx,\beta)=z$ where
$z$ is an element of ${\mathbb Z}_N$; i.e.,
$z \in ( 1, e^{2\pi i/N} , e^{4\pi i/N}, \cdots, e^{2(N-1) \pi i/N }) \times \mathbf{1}$
where $\mathbf{1}$ is the $N\times N$ identity matrix.

Such a curve in $SU(N)$ cannot, in general,
become a constant in $t$ as $|\vec{x}| \rightarrow \infty$ in $g(\vec{x},t)$ since
the end points of the curve are fixed to be 
$\mathbf{1}$ and $z \neq {\mathbf 1}$. 
Hence
\begin{equation}
\lim_{|\vec{x}| \rightarrow \infty} \partial_0 g(\vec{x},t) \neq 0 \quad {\rm identically}
\end{equation}
Thus, even if we start with $A_0^\infty = 0$, a gauge transformation by
such an element can lead to a nonzero $A_0^\infty =  g^{-1} \partial_0 g\bigr]_{\vert\vx\vert \rightarrow \infty} \neq 0$.
It is well known that such gauge transformations are important in finite temperature field theory with the Polyakov loop
\begin{equation}
L={\rm Tr}\, \exp \left(i \int_0^\beta dt A_0(\vec{x},t) \right)
\label{6}
\end{equation}
considered as an order parameter.
There are also indications from numerical work
that the expectation value of $L$ need not be the identity at high temperatures, 
consistent with the possibility that $A_0^\infty$ can be nonzero.

Our second example is a classic one, the Josephson effect. The basic setup here is that one has
a normal conductor sandwiched between two superconductors.
The normal material constitutes a spacetime region with disjoint boundaries.
Thus
the electrons in the normal material can be described by electrodynamics for a region
with boundaries. Generally we can have different phases for the many-body wave function
in the two superconducting regions. The phases of the electron at the boundaries of
the normal
material have to be matched to these. The difference of phases leads to the Josephson
current, with the time-derivative of the phase acting as the value of
$A_0$ on the boundary. This is another case of a gauge theory in a physical context
 with a nonzero values
of $A_0$ on the boundaries.

\section{Physical implications of boundary data}

We now consider some physical implications of nontrivial boundary data.
We will consider two examples: a field theory defined on a manifold with disjoint
boundaries, and chemical potentials for nonabelian charges in the presence of
magnetic monopoles.
\subsection{Manifolds with disjoint boundaries}

The simplest case would be to start with a nonabelian gauge theory, say, QCD, 
in 1+ 1 dimensions, with the real line as the spatial manifold.
 In this case we have disjoint boundaries corresponding to
 $\vert\vx \vert \rightarrow \pm \infty$.
 Let $A_0^{\pm\infty}$ denote the two boundary values of $A_0$; i.e., 
 $A_0 \rightarrow A_0^{\pm\infty}$ as $\vert \vx \vert \rightarrow \pm\infty$,
 respectively. We can then decompose $A_0$ as 
 \beqar
 A_0 &=& A_0^{+} + A_0^{-}\nonumber\\
 A_0^\pm &\rightarrow& A_0^{\pm\infty} ~~{\rm as} ~~\vert\vx \vert \rightarrow \pm \infty \nonumber\\
  A_0^\pm &\rightarrow &0 ~~{\rm as} ~~\vert\vx \vert \rightarrow \mp \infty
 \label{7}
 \eeqar
 This is essentially a decomposition of $A_0$ via a resolution of unity over the real line. 
 The specific choice of $A_0^\pm$ is not important, different choices with the same boundary values
 differ by terms proportional to the Gauss law constraint.
 In what follows, we will, for simplicity, consider the case with $\del_0 A_0^\pm \rightarrow 0$
 as $\vert\vx \vert \rightarrow \pm \infty$; i.e., the boundary values of $A_0$ are taken to be independent
 of time. For physical states $\ket{\rm phys}$ which obey the Gauss law, we have
 \beqar
 Q(A_0) &=& Q(A_0^+ ) + Q(A_0^- )\nonumber\\
{} [Q(A_0^+ ), Q(A_0^- ) ] \, \ket{\rm phys} &=& 0
 \label{8}
 \eeqar
 Effectively we have two operators $Q(A_0^\pm )$ associated  to charges defined via fluxes on the two
 disjoint boundaries.
 Since the Hamiltonian has a term $Q(A_0)$, once a choice of $A_0^{\pm\infty}$ has been made, the residual symmetry of the theory
 is given by the common stability group of  $Q(A_0^\pm )$. In general, the stability groups for these
 two operators $Q(A_0^\pm )$ can be different.
 Transformations corresponding to the charge algebra, with parameters
 $\theta$, are generated by $Q(\theta ) = \int \Tr (E\, D \theta )$. Analogous to how $A_0$ was decomposed,
 we can write
  \beqar
\theta &=& \theta^{+} +  \theta^{-}\nonumber\\
  \theta^\pm &\rightarrow&  \theta^{\pm\infty} ~~{\rm as} ~~\vert\vx \vert \rightarrow \pm \infty \nonumber\\
   \theta^\pm &\rightarrow &0 ~~{\rm as} ~~\vert\vx \vert \rightarrow \mp \infty
 \label{9}
 \eeqar
This leads to two copies of the algebra of charges, corresponding to
$Q(\theta^+)$ and $Q(\theta^-)$.
 
As mentioned above, the boundary values and hence the stability groups of
$Q(A_0^\pm\infty)$ can be different for different disjoint components of the boundary.
Two examples of this for $SU(3)$ would be
the choices
\beqar
A_0^{+\infty} = a_0^{+\infty} {\rm diag} (1,1,-2), &\quad& A_0^{-\infty} = a_0^{-\infty} {\rm diag} (1,-2,1) 
\nonumber\\
A_0^{+\infty} = a_0^{+\infty} {\rm diag} (1,1,-2), &\quad&
A_0^{-\infty}=b_0^{-\infty} {\rm diag}(1,-1,0) \label{bc2}
\eeqar

One of the physical consequences of such unequal choices on disjoint boundaries is that
it can lead to a current. This is similar to how a tunneling current arises
in the Josephson junction mentioned earlier.
To demonstrate how this can happen, we consider the theory with the addition of fields which
carry nonzero charges. For simplicity, we take them to be Dirac fields, so that the
action written out in first order form is
\beqar
S &=& \int \Bigl[ E_i^a {\dot A}_i^a - {1\over 2} ( E_i^a E_i^a + B^a_i B^a_i )
- E_i^a (D_i A_0)^a + i {\bar \psi}  \gamma_0 \del_0  \psi
- i {\bar \psi}  \gamma_i \del_i \psi \nonumber\\
&&\hskip .2in + i {\bar \psi}  \gamma_0 A_0 \psi
- i {\bar \psi} \gamma_i A_i \psi \Bigr]
\label{10}
\eeqar
We can now write $A_0 = g^{-1} {\dot g}$ for some group element
$g$ which is constant on each connected piece of the boundary
at fixed $t$. Using this we can rewrite the action
(\ref{10}) as
\beqar
S &=& \int \Bigl[ \E_i^a {\dot \A}_i^a - {1\over 2} ( \E_i^a \E_i^a + \B^a_i \B^a_i )
 + i {\bar \psi}  \gamma_0 \del_0  \psi
- i {\bar \psi}  \gamma_i \del_i \psi \nonumber\\
&&\hskip .2in + i {\bar \psi}  \gamma_0 g^{-1}{\dot g} \psi
- i {\bar \psi} \gamma_i (g^{-1} \A_i  g + g^{-1} \del_i g) \psi \Bigr]
\label{11}
\eeqar
where
\beq
\A_i = g \, A_i \, g^{-1} - \del_i g \, g^{-1}
\label{12}
\eeq
with $\E_i^a$ as its canonical conjugate, and $\B$ and $\D$ are constructed using
$\A$. We can take $\A_i$ and $\E_i$ as the basic phase space variables for the gauge field.
For this choice of variables, the $g$-dependence is transferred entirely to the matter terms.
The dependence of the partition function or the functional integral
on the boundary values can be obtained by varying $g$.
With $g^{-1}\delta g = -i t^a \delta\theta^a$, we find
\beqar
\delta S 
&=& \delta S_{\rm surf} +\int \left[ - (D_0 J_0)^a+ (D_i J_i )^a  \right] \, \delta\theta^a\, ,
\nonumber\\
\delta S_{\rm surf} &=& \int_V {\bar \psi} \gamma_0 t^a \psi \, \delta \theta^a\Bigr]^{t_f}_{t_i}
- \int_{\del V}{\bar \psi} \gamma_i t^a\psi \,\delta \theta^a dS_i \, dt \, ,\label{13}\\
J_0^a &=& {\bar\psi} \gamma_0 t^a \psi, 
\quad J_i^a = {\bar \psi} \gamma_i t^a \psi \, .
\nonumber
\eeqar
We are considering the fields in a spatial volume $V$ with the time interval
as $[t_f, t_i]$.
The boundary $\del V$ can have disjoint components.
The last set of terms in (\ref{13}) will vanish in the classical case
by current conservation.
In the quantum theory, the conservation law, which follows from variation of the action
can be imposed as an operator equation inside matrix elements or
at the level of the integrand of the functional integral.
To see this in more detail, consider the vacuum-to-vacuum transition amplitude
for a set of fields generically denoted as $\vf$,
\beq
Z\equiv \braket{0|0}= \int [d\vf] \, \Psi_0^* [\vf_f, t_f] ~ e^{i S (\vf, t_f, t_i )}\, \Psi_0 [\vf_i, t_i]
\label{13a}
\eeq
where $S$ is the action evaluated for field configurations
$\vf(\vx , t)$ with $\vf(\vx, t_f) = \vf_f$, $\vf(\vx, t_i) = \vf_i$.
We integrate over $\vf_f, \vf_i$ as well with the wave functions as the weighting factors.
Consider the functional derivative with respect to the bulk values of the
fields, which keep $\vf_f, \vf_i$ and the spatial boundary value
of $\vf(\vx, t)$ fixed.
The vanishing of the integral of a total derivative then leads to the identity
\beq
0 = \int [d\vf]  \Psi_0^* [\vf_f, t_f]~ \left[ {\delta \over \delta \vf} e^{iS} \right]
\Psi_0[\vf_i, t_i]
= i \int [d\vf]  \Psi_0^* [\vf_f, t_f]~ e^{iS} \, {\delta S \over \delta \vf} \, \Psi_0[\vf_i, t_i]\,.
\label{14}
\eeq
(If we have sources, there will be additional terms corresponding to their transformation; this is standard
procedure for the derivation of Ward-Takahashi identities.)
This shows how the bulk equations of motion can be realized in matrix elements.
For the more general variation of the vacuum-to-vacuum amplitude with
no sources, we can then write
\beqar
Z' = Z +
\int \delta \theta^a \, {\delta Z \over \delta \theta^a}
&=&\int [d\vf]\,\Psi_0^* [\vf_f +\delta \vf_f, t_f]~ e^{i (S + \delta S)}
~\Psi_0 [\vf_i+ \delta \vf_i, t_i] \nonumber\\
&=& \int [d\vf]\, \Psi_0^* [\vf_f +\delta \vf_f, t_f]~ e^{iS} \, e^{ i \delta S_{\rm surf}}
~\Psi_0 [\vf_i+ \delta \vf_i, t_i] 
\label{15}
\eeqar
The surface terms in $\delta S_{\rm surf}$
on the two time-slices at $t_f$ and $t_i$ can be absorbed into
the change of the final and initial wave functions, since the
wave functions transform as
$\delta \Psi(\vf, t) = i \bigl[\delta S_{\rm surf}\bigr]_{{\rm at} \, t} \, \Psi (\vf, t)$.
Thus, upon using
(\ref{13}), (\ref{14}),
we can simplify (\ref{15})
as
\beq
\int \delta \theta^a \, {\delta Z \over \delta \theta^a}
= - i \int  [d\vf] \, \Psi_0^* [\vf_f, t_f]~e^{iS} \, \left[ \sum F^a_\alpha \, \delta \theta_\alpha^a\right]
~\Psi_0(\vf_i, t_i)
\label{16}
\eeq
where $F^a_\alpha = \int_{{\del V}_\alpha}  J_i^a dS_i$ is the flux of the current across
each connected component (indexed by $\alpha$) of the spatial boundary.
These arguments go through in Euclidean signature with $Z$ being the
partition function.

Now consider calculating the partition function (or the vacuum-to-vacuum amplitude)
$Z$ using the action
as given in (\ref{10}).
 The result will obviously depend on the boundary values of $A_0$,
or, equivalently, on $g$. So the variation of $Z$ with respect to
$g$ is, in general, not zero.
The second way of calculating involves using the action
(\ref{11}). In this case, the dependence of $Z$ on $g$ is given by
the expectation value of the current fluxes as in (\ref{16}).
The compatibility of these two ways of calculating $Z$ tells us
that
if the functional integral or partition function depends on the
boundary values of $A_0$, then there must be current fluxes across
the boundaries.
Notice that, with disjoint boundaries, we can vary $g$ independently
on the different connected components, and so the individual fluxes
cannot be zero.

Although we started with the $1+1$ dimensional case, this part of the discussion
(starting with (\ref{10}) ) is general
and applies to theories in higher dimensional manifolds with disjoint boundaries.
To summarize, if we have different values of $A_0$ on different disconnected
pieces of the boundary, there will be current fluxes in the bulk of spatial volume.
We also note that if the operators which commute with
the boundary data form a nonabelian algebra, in the quantum theory,
they are associated with a
multiplicity of edge states. As these  commutants can differ at the different edges,
these multiplicities also can differ.

We have already remarked on the similarity of what we do here and the tunneling current
in a Josephson junction. A physical situation where our arguments for the nonabelian case 
would apply
is for neutron stars. The general expectation is that
in the interior of a neutron star there are concentric layers of hadronic matter in different phases
\cite{glendenning}. These include pure quark phase, quark slabs, etc. The interfaces of
such regions function as the disjoint boundaries of our discussion above.
It is difficult to make more specific predictions at this stage, since the
physics of the hadronic phases in the neutron star is still 
not well understood.

\subsection{Magnetic monopoles}

In the previous example, we considered $A_0$ to be a constant on the boundary.
But there can be topological obstructions to this choice. The simplest example which can illustrate
this is the case of a background corresponding to
 the 't Hooft-Polyakov monopole.
 Recall that the Higgs field $\Phi$ in this case has the asymptotic behavior
\begin{equation}
\Phi_{|\vec{x}| \rightarrow \infty} \rightarrow \Phi^\infty(\hat{x}) = f_0^\infty\, \vec{\tau} \cdot \hat{\phi}(\hat{x}) \label{aPhi0}
\end{equation}
where the winding number of the map $\hat{x} \rightarrow \hat{\phi}(\hat{x})$ is the monopole charge
and $f_0^\infty$ is a constant. ($\tau^a$ are the Pauli matrices, viewed as the generators
of the Lie algebra of $SU(2)$.)

The Hamiltonian (\ref{Ham2}) has the term $Q(A_0)$ in addition to the usual bulk terms.
The boundary value of $A_0$ must be such that the Hamiltonian commutes with the
Higgs field, so that the value of the Higgs field at spatial infinity is not changed
by time-evolution.
This is needed for the chosen field configuration to correspond to a 
sector with spontaneously broken symmetry (and a static monopole field).
Specifically, we must require that
\begin{equation}
[Q(A_0), \Phi]_{\, |\vec{x}|\rightarrow \infty} \rightarrow 0
\label{17}
\end{equation}
If $A_0^\infty(\hat{x})$ is the boundary value, i.e., 
$A_0(\vec{x})_{|\vec{x}|\rightarrow \infty} \rightarrow A_0^\infty(\hat{x})$,
the requirement (\ref{17}) leads to
\begin{equation}
A_0^\infty(\hat{x}) =a_0^\infty\,  \vec{\tau} \cdot \hat{\phi}(\hat{x})\label{bcA}
\end{equation}
where $a_0^\infty$ is a constant.
Thus we cannot choose $A_0$ to be a constant on the two-sphere at
spatial infinity, it depends on $\vec{x}$ as indicated. 
$A_0^\infty(\hat{x})$ may be viewed as the chemical potential for nonabelian
charges in the presence of the monopole. The term in the
Hamiltonian to be used to define the partition function for a system with finite
charge is thus
\begin{equation}
Q(A_0) = \int d^3 x\, {\rm Tr} \,(E_i D_i A_0 ) \label{chempot}
\end{equation}
with $A_0^\infty(\hat{x})$ as in (\ref{bcA}).
On quantum states, because of the Gauss law,
$Q(A_0)$ depends only on the equivalence class $[A_0]$. Thus
 if $A_0^{(i)} \in [A_0]$, then $(A_0^{(1)} - A_0^{(2)})_{|\vec{x}| \rightarrow \infty} \rightarrow 0$.
 
An additional feature or nuance associated to this
can be seen by considering
the case of the grand unified $SU(5)$ theory
 \cite{gutsu5}, broken down, as usual, to the standard model gauge group.
In this case, the Higgs field $\Phi$ belongs to the representation
\underline{24} of $SU(5)$;
this may be viewed a traceless $5 \times 5$ matrix. 
In the non-monopole sector, it breaks $SU(5)$ to $H = \bigl[ SU(3) \times SU(2) \times U(1)\bigr] / \mathbb{Z}_6$; its asymptotic value on the two-sphere at spatial
infinity is of the form
\begin{equation}
\Phi^\infty = \phi^\infty \,{\rm diag}\left(1,1,1,-{\textstyle {3\over 2}},-{\textstyle{3\over 2}}\right)
\label{18}
\end{equation}
where
$\phi^\infty$ is a constant.
The $5^*$ matter multiplet is $(d_1^c, d_2^c, d_3^c, e^-, \nu_e)_L$. Thus, we see clearly that
(\ref{18}) leaves color rotations of the three $d^c$'s invariant, and similarly for the
$SU(2)$ rotations on $(e^-, \nu_e)$.
Defining $ T_3 = {\half} {\rm diag} (0,0,1,-1,0)$,
we can write the asymptotic value of the field as
\begin{equation}
\Phi^\infty = \phi^\infty \left[\frac{5}{2} T_3 + {\rm diag} (1,1,- {\textstyle{1\over 4}},-{\textstyle{1\over 4}},-{\textstyle{3\over 2}}) \right] \label{aPhi1}
\end{equation}
The reason for writing it this way is that we can regard the $SU(5)$ monopole as
the 't Hooft-Polyakov monopole
of an $SU(2)$ subgroup acting on, say, $(d_3^c, e^-)$ broken down to
a $U(1)$. Thus, for a
generic monopole ansatz, the asymptotic value
of the Higgs field can be obtained by rotating (\ref{aPhi1})over the sphere $S_\infty^2$ at infinity. In the lowest monopole sector, it is given by
\begin{equation}
\Phi^\infty(\hat{x}) = \phi^\infty \left[\frac{5}{2} T \cdot \hat{x} + {\rm diag} (1,1,- {\textstyle{1\over 4}},-{\textstyle{1\over 4}},-{\textstyle{3\over 2}}) \right] \label{aPhi2}
\end{equation}
For a general winding number solution, we can replace $\hat{x}$ by $\hat{\phi}(\hat{x})$ in (\ref{aPhi2}) where $\hat{\phi}(\hat{x})$ is as in (\ref{aPhi0}).

The $SU(5)$ monopoles are nonabelian. From the structure of the spontaneous symmetry breaking, 
it would seem that one can introduce color $SU(3)$ rotations and $SU(2) \times U(1)$ rotations
as collective coordinates or moduli whose quantization would convert the classical solution
into a suitable multiplet of $H = \bigl[ SU(3) \times SU(2) \times U(1)\bigr] / \mathbb{Z}_6$. 
However, it is known that this cannot be done
\cite{Balachandran:1982gt}.
There is a topological obstruction, and typically, in the monopole sector, the 
unbroken subgroup $H'$ is smaller than $H$ and depends on the choice of
the monopole solution.
For the asymptotic behavior given in
(\ref{aPhi2}), the unbroken group is $ H' = U(2) \times U(1) \times U(1)$. 
The full theory is discussed in \cite{Balachandran:1983xz,
Nelson:1983fn}. 

Now consider a gas of gluons, $W$'s and other particles carrying
nonabelian charges
in the presence of a monopole characterized by, say, (\ref{aPhi2}).
For a statistical distribution of these particles
with nonzero charge, we need the chemical potential.
However, as mentioned above, it
must commute with the Higgs field of the monopole.
 So $A_0^\infty$, to be used as the chemical potential,
 can be any element in the Lie algebra of $U(2) \times U(1) \times U(1)$.

The older work \cite{Balachandran:1982gt}-\cite{Nelson:1983fn} shows the topological obstruction
to color and $SU(2)$ rotations of the monopole.
The conclusion of the
present argument is that this breakdown of color in the presence of
such monopoles applies also to
forming statistical distributions of the charged particles on this background.

Regarding the statistical distributions of particles, it may also be interesting to note that
some of the particles can undergo a change of statistics.
Notice that the Higgs field in (\ref{aPhi2}) is invariant under a simultaneous
rotation of ${\hat{x}}_i$ and
$T_i$. This means that the angular momentum operator in this background
will be a combination $L_i + S_i +  T_i $, where $L_i$ and $S_i$ denote the
orbital and spin angular momentum operators, respectively. This is the 
well known ``spin from isospin" in a monopole background \cite{jackiw}.
For the present context this means that the particles corresponding to
 $(d_3^c, e^-)$ will change statistics. 
 Many of the gauge bosons become massive at the scale of $SU(5)$ breaking.
 The gauge particles which remain massless at this stage will retain their
 statistics since they transform as the adjoint representation of the charge algebra and
 hence the contributions from $T_i$ will be integer-valued.

\subsection{The realization of the charge algebra}

We will now make some comments on how the charge operators act on the states,
i.e., on the nature of the representation of the charge algebra.
These statements are based on the Fabri-Picasso \cite{Fabri:1966} theorem and are basically
the application of that result to the present context. 

We note however that these theorems require the existence of a mass gap separating vacuum 
and the excited states  \cite{orzalesi}.
There is no rigorous proof that there is such a mass gap in nonabelian  gauge theories.

Consider the charge operators
given in (\ref{4}). The Hamiltonian has a term $Q(A_0)$ and we have
the commutation rule $[Q(\mu), \H ] = -i \,Q([\mu, A_0])$. By the Gauss law,
only the boundary value of the commutator $[\mu, A_0]$ is relevant when this
relation acts on physical states.
Thus the commutant of $A_0^\infty$ in the Lie algebra
defines the symmetry group of the problem, which we will
refer to as $G_{A_0}$ for brevity.

The momentum operator $P_i$ is a crucial ingredient for the
Fabri-Picasso theorem, so it is useful to consider the expression for it
in our case.
Explicitly,
\beq
P_i = \int d^3x\, E^a_j \, F_{ij}^a
\label{19}
\eeq
If $\xi_i$ is a constant vector, we can write
\beq
\xi_i P_i = \int d^3x\, E^a_i (\xi\cdot \del) A^a_i - \int d^3x\, E^a_i D_i(A\cdot \xi )
\label{20}
\eeq
The last term will generate a gauge transformation with parameter $A\cdot \xi$, if
we take $A_i$ to vanish at the spatial boundary. It is then clear from the 
first term that $P_i$ generates translations on all gauge-invariant quantities
or on operators acting on physical states. From the commutation rule (\ref{CCR}),
we can easily check that
\beq
[ Q(\mu ), P_i ] = 0
\label{21}
\eeq
It should be emphasized that this relation holds even for
functions $\mu$ which do not vanish on the spatial boundary or spatial
infinity. No integration by parts or discarding of boundary terms is needed.

We also note that the charge operator $Q(\mu)$, with constant value for $\mu$ at the boundary,
can be written in terms of a density
using the Gauss law. On physical states, we can write
\beq
Q(\mu ) \ket{\rm phys} = \oint \mu^a \, E^a_i \, dS_i ~\ket{\rm phys}
\label{22}
\eeq
Integrating $\mu^a \left( \nabla \cdot E^a + f^{abc} A_i^b E^c_i \right) \, \ket{\rm phys} = 0$
over all space, taking $\mu^a$ to be a constant equal to the value on the boundary, we find
\beq
Q(\mu ) \ket{\rm phys} = -  \int d^3x\, \left(\mu^a f^{abc} A_i^b E^c_i \right)\, \ket{\rm phys}
\equiv \int d^3x\, J_0 (x) \, \ket{\rm phys}
\label{23}
\eeq
We can now see the Fabri-Picasso theorem for this case as follows.
\beq
\bra{0} J_0 (x) \, Q(\mu) \ket{0} =  \bra{0} e^{i P\cdot x} \, J_0 (0) \, e^{-iP \cdot x} \, Q(\mu )
\ket{0} = {\rm constant}
\label{24}
\eeq
using (\ref{21}) and $P_i \ket{0} = 0$.
This immediately gives
\beqar
\bra{0} Q(\mu ) Q(\mu ) \ket{0} &=& \int d^3x\, \bra{0} 
J_0 (0) \,Q(\mu) \ket{0} = {\rm constant} \times \int d^3x\nonumber\\
&=& {\rm constant}\times \infty
\label{25}
\eeqar
This result shows that either $Q(\mu ) \ket{0} = 0$ (corresponding to the vanishing of the constant
in this equation) or
the state $Q(\mu ) \ket{0}$ does not exist as it is not normalizable.

So far, this discussion is similar to what happens with spontaneous 
symmetry breaking.
But for us, there is an additional condition because the Hamiltonian
contains a term which is $Q(A_0)$. Thus $H \ket{0}$ produces a
term $Q(A_0 ) \ket{0}$ and if this is not normalizable, the vacuum state
will not be in the domain of the Hamiltonian. So the only physically sensible
choice is
$Q(A_0 ) \ket {0} = 0$. Notice that this also tells us that the vacuum energy is not shifted
by the presence of the term $Q(A_0)$.
Now consider those $Q(\mu)$ which weakly commute with $Q(A_0)$, with $\mu \neq A_0$.
These elements, along with $Q(A_0)$ generate
the subgroup $G_{A_0}$. For these $Q(\mu )$, there is no real restriction, 
either choice ($Q(\mu) \ket{0} = 0$ or $Q(\mu ) \ket{0} \neq 0$) is possible.
Thus we can have a situation where $G_{A_0}$ is fully realized as a symmetry, or
partially or fully broken spontaneously to the $U(1)$ group generated by
$Q(A_0)$.

Now let us turn
 to the choices of $\mu$ with $[Q(\mu), Q(A_0)] = -i \,Q([\mu, A_0]) \neq 0$.
 These would generate elements of the gauge group $G$ which are not 
 contained in $G_{A_0}$ and hence are not symmetries of the theory.
 Again, there are two possibilities: either $Q(\mu ) \ket{0} = 0$ 
 (with $[\mu, A^\infty_0] \neq 0$)
 or its action on $\ket{0}$ is not defined following the argument given above.
Coleman's theorem \cite{Coleman:1966} would suggest that 
the case $Q(\mu ) \ket{0} = 0$ can lead to additional symmetries
and hence should be ruled out
since all $Q(\mu)$ which commute with
 the Hamiltonian have been included as generators of
 $G_{A_0}$. 
 But it is not clear that the premises of
 Coleman's theorem (which inlcude
 Lorentz invariance and the Reeh-Schlieder theorem)
 are obtained in our case.
 
 To summarize, there is always an unbroken $U(1)$ subgroup generated by
 $Q(A_0)$; one may have a larger symmetry $G_{A_0}$, which may be partially
 or fully spontaneously broken down to this $U(1)$. 
 (Since we are focusing only on the Yang-Mills action, the explicit mechanism for such a
 breaking is not manifest; we would have to include additional charged
fields to analyze this.)
 Transformations corresponding to elements of
 the global group $G$ which are not in
 $G_{A_0}$ are not symmetries of the theory.
 
Another point of interest is that the term $Q(A_0)$ in the Hamiltonian can lead to 
 energy corrections to the charged states.
Writing $\H = \H_0 + Q(A_0)$, notice that $\H_0$ and $Q(A_0)$ commute with each other 
and can be
simultaneously diagonalized. A nonzero $A_0^\infty$ can lead to splitting of 
degenerate states for the charged particles like gluons, charge being
defined by $Q(A_0)$.
We also note that while there are many issues associated with the implementation of
nonabelian charge rotations on the charged sector in general \cite{Balachandran:2014voa},
in the present case, we are only dealing with an Abelian group defined by
$Q(A_0)$.
 
 \bigskip

This research was supported in part by the U.S.\ National Science
Foundation grant PHY-1820721
and by PSC-CUNY awards.
 
%%%%%%%%%%%%%%%%%%%%%%%%%%%%%%%%%%%%%%%%%%%%%%%
%%%%%%%%%%%%%%%%%%%%%%%%%%%%%%%%%%%%%%%%%%%%%%%

%%%%%%%%%%%%%%%%%%%%%%%%%%%%%%%%%%%%%%%%%%%%%%%
%%%%%%%%%%%%%%%%%%%%%%%%%%%%%%%%%%%%%%%%%%%%%%%
%%%%%%%%%%%%%%%%%%%%%%%%%%%%%%%%%%%%%%%%%%%%%%%
%%%%%%%%%%%%%%%%%%%%%%%%%%%%%%%%%%%%%%%%%%%%%%%
\end{document}